\documentclass[preprint,12pt]{elsarticle}

\usepackage{amssymb}
\usepackage{amsmath}
\usepackage{amsthm}
\usepackage{algorithm}
\usepackage{algpseudocode}
\usepackage{tabularx}
\usepackage{graphicx}
\usepackage{subcaption}
\usepackage{todonotes}

\journal{Computer}

\begin{document}

\begin{frontmatter}


\title{Resilient Packet Forwarding: A Reinforcement Learning Approach to Routing in Gaussian Interconnected Networks with Clustered Faults}

\author[first]{Mohammad Walid Charrwi}
\author[second]{Zaid Hussain}

\affiliation[first,second]{organization={High Performance Computing Lab, Computer Science Department, Kuwait University},country={Kuwait}}

\begin{abstract}
As Network-on-Chip (NoC) and Wireless Sensor Network architectures continue to scale, the topology of the underlying network becomes a critical factor in performance. Gaussian Interconnected Networks based on the arithmetic of Gaussian integers, offer attractive properties regarding diameter and symmetry. Despite their attractive theoretical properties, adaptive routing techniques in these networks are vulnerable to node and link faults, leading to rapid degradation in communication reliability. Node failures (particularly those following Gaussian distributions, such as thermal hotspots or physical damage clusters) pose severe challenges to traditional deterministic routing. This paper proposes a fault-aware Reinforcement Learning (RL) routing scheme tailored for Gaussian Interconnected Networks. By utilizing a PPO (Proximal Policy Optimization) agent with a specific reward structure designed to penalize fault proximity, the system dynamically learns to bypass faulty regions. We compare our proposed RL-based routing protocol against a greedy adaptive shortest-path routing algorithm. Experimental results demonstrate that the RL agent significantly outperforms the adaptive routing sustaining a Packet Delivery Ratio (PDR) of $\approx$0.95 at 40\% fault density compared to $\approx$0.66 for the greedy. Furthermore, the RL approach exhibits effective delivery rates compared to the greedy adaptive routing, particularly under low network load of 20\% at $\approx$ 0.57 vs. $\approx$ 0.43, showing greater proficiency in managing congestion, validating its efficacy in stochastic, fault-prone topologies.
\end{abstract}

\begin{keyword}
Distributed Computing \sep Faulty Nodes \sep Greedy Adaptive Routing \sep Communication \sep Reinforcement Learning.
\end{keyword}

\end{frontmatter}

\section{Introduction}
\label{section:introduction}
Interconnection networks form the backbone of modern multiprocessor systems, many-core architectures, and network-on-chip designs. The continuous scaling of semiconductor technologies has facilitated the creation of massive multi-core systems and large-scale sensor networks \cite{benini2002networks}. In these environments, fault adaptivity is not merely a feature but a primary design constraint. The reliability of packet delivery in wireless mesh and sensor networks is crucial, especially in environments subject to frequent node failures. Traditional routing algorithms, often relying on global knowledge or simple, static metrics (like shortest path), fail catastrophically when a significant number of network nodes become faulty \cite{duato2003interconnection}. This challenge necessitates the adoption of fault-adaptive routing. In realistic operational scenarios, faults rarely occur in a uniform, random distribution. Instead, they frequently follow a Gaussian distribution, where a "seed" failure (caused by localized overheating, battery depletion, or physical damage) increases the probability of neighboring failures \cite{skadron2004temperature}. This phenomenon creates "fault clusters" that sever connectivity in standard mesh topologies. Traditional routing algorithms rely on fixed paths or slow-converging updates. When a significant fault block occurs, these algorithms often suffer from local minima, high packet loss, or deadlock \cite{ebda}. While adaptive routing algorithms exist, they often struggle with the complex, non-Cartesian links found in more advanced topologies like Gaussian Interconnected Networks \cite{elia2003class}. Gaussian Interconnection Networks are constructed over the ring of Gaussian integers \cite{huber1994codes} and provide rich structural properties, including high symmetry, low diameter, and strong theoretical connectivity. These characteristics make Gaussian Interconnected Networks an attractive target for scalable communication systems. However, practical deployment requires routing strategies that can adapt to unpredictable faults and dynamically changing network conditions. 

Reinforcement learning has emerged as a powerful paradigm for sequential decision-making in complex environments. By learning from interaction, RL agents can adapt policies that outperform static heuristics in dynamic and uncertain scenarios \cite{sutton2018reinforcement, kaelbling1996reinforcement}. Recent advances in deep reinforcement learning have enabled scalable learning in high-dimensional state spaces across different domain fields such as hardware design \cite{mirhoseini2021chip}, cybersecurity \cite{Nguyen_2023}, quantum computing \cite{charrwi2025tpu, rlquantum} along with interconnected networks \cite{charrwi2025selfhealingnetworksonchiprldrivenrouting}. This makes RL feasible to be applied to routing problems. This work explores the integration of RL into Gaussian network routing and evaluates its benefits over traditional methods.

The objective of this research is to rigorously quantify the performance limitations of the Greedy Adaptive Routing protocol when faced with increasing fault density in a Gaussian Interconnected network and to demonstrate the performance benefits achieved by RL-based routing policy designed to overcome these limitations. The core contribution of this work lies in:

\begin{itemize}
    \item Designing a hybrid, multi-objective reward function for the RL agent that concurrently optimizes for path efficiency (via step cost)
    \item Providing a detailed comparative analysis using the key metrics of Packet Delivery Ratio (PDR), Adaptive Score (PDR vs. Fault Density), and Normalized Throughput, confirming the RL agent's ability to maintain network efficiency and reliability under fault load.
\end{itemize}

The paper is organized into different sections. Section \ref{section:relatedwork} is the related work sections which discusses the relevant work done in the field of routing. After that, Section \ref{section:background} discusses the background and relevant definitions that will be used in this paper. Next, Section \ref{section:methadology} is the methodology section which explains the approach for the implementation of the greedy adaptive routing. Furthermore, it also formulates the reward representation for the RL-based approach of routing. In Section \ref{section:experimental} we present a series of experimental evaluations by carrying out different experiments for greedy adaptive routing against the RL-based routing. We also quantify different metrics performance of the network under different fault and load capacity for the Gaussian Interconnected Network. Finally Section \ref{section:conclusion} concludes the paper by deriving the quantified improvement in the RL-based routing over the greedy adaptive routing, and highlights proposed future works.

\section{Related Work}
\label{section:relatedwork}
\subsection{Foundation of Structured Network Routing and Fault Adaptivity}
Structured interconnection networks, including the $k$-ary $n$-cube and mesh networks \cite{dally2004principles}, are designed for predictable and high-speed communication. Gaussian networks, defined by the set of Gaussian integers $\mathbb{Z}[i]$ (a two-dimensional lattice where $\mathbb{Z}$ represents the set of all standard integers and i is the imaginary unit $\sqrt{-1}$) modulo for any $\alpha \in \mathbb{Z}[i]$ (where $\alpha$ is a specific Gaussian integer used as a "modulo" to define the network's boundaries and wrap-around structure), present an algebraic topology with highly regular properties, including excellent distance distribution and decomposability into Hamiltonian cycles. Initial work in these networks focused on finding optimal shortest path routing algorithms in fault-free environments \cite{HDGaussian}. The challenge of fault adaptivity for routing in these structured topologies has led to the development of several adaptive techniques. Many conventional fault-adaptive algorithms rely on explicit mechanisms for fault detection and information dissemination, such as broadcast or specific fault notification messages, to update the network's known faulty components and recalculate paths \cite{tripathiFaultTolerantRouting}. The focus remains on guaranteeing path existence and avoiding deadlocks, often by enforcing specific turn restrictions or virtual channel usage, particularly in NoC environments \cite{kurokawa2025adaptive}. 

\subsection{Reinforcement Learning for Adaptive Policy Generation}
Greedy adaptive routing (forward to the neighbor reducing distance while avoiding faulty links) is simple to implement but can fail to find routes when local minima occur or faults block the obvious path. In contrast, RL-based methods can learn alternate routes through experience. This motivates our comparison: we implement a greedy adaptive router and a RL agent in a Gaussian Interconnected topology, building on prior insights \cite{boyan1993packet, wu2023rlara} to quantify how RL improves fault adaptivity and throughput. The paradigm shift towards Reinforcement Learning (RL) in network control has been motivated by the need for protocols that can dynamically and autonomously adapt to changing environments without predefined rules. The efficacy of an RL routing agent is highly dependent on the design of its reward function \cite{goto2025design}. Reward functions can be broadly categorized as purpose-oriented (rewards granted only upon goal achievement) or process-oriented (rewards or penalties granted for intermediate states and actions) \cite{goto2025design}. While purely purpose-oriented rewards can achieve the goal, they may result in highly inefficient or unnecessarily long paths. Modern RL-based routing protocols, therefore, often employ a hybrid approach to achieve multi-objective optimization, integrating cost factors such as latency, congestion, or path length into the reward structure alongside the final destination reward \cite{PEDTARA}.

\subsection{Performance Metrics for Protocol Assessment}
The performance of fault-adaptive routing protocols is typically assessed using a combination of reliability and efficiency metrics.
\begin{itemize}
    \item Packet Delivery Ratio (PDR): PDR is the fundamental measure of reliability, calculated as the ratio of successfully received packets to the number of packets transmitted. When PDR is evaluated as a function of environmental change (e.g., fault density), it serves as the Adaptive Score, quantifying the protocol's resilience to topology degradation. 
    \item Normalized Throughput: Throughput provides a measure of the network's capacity and efficiency, inherently capturing the impact of path length and congestion. Protocols that find shorter paths and avoid congested or faulty regions generally achieve higher throughput, as the effective delivery rate is inversely correlated with the load and the path length in congested networks.
\end{itemize}


\section{Background}
\label{section:background}
To analyze the topological structure and properties of interconnected networks, they can be abstracted as graphical models within a two-dimensional domain. This approach employs concepts from graph theory to represent network layouts and characteristics, thereby facilitating a systematic examination of their connectivity and structural features. Vertices in in these graphical models represent nodes and edges denote communications links between these nodes. A graph \( G = (V, E) \) is formally defined as an ordered pair comprising a vertex set \( V(G) \) and an edge set \( E(G) \). Each edge \( (v, u) \in E \) is undirected and connects a pair of distinct vertices \( v, u \in V \). The graph \( G \) is considered \textit{regular} if every vertex shares the same \textit{degree}, which denotes the number of edges incident to that vertex. A \textit{path} \( P(a, b) \) from vertex \( a \in V \) to vertex \( b \in V \) is a sequence of distinct vertices \( v_1, v_2, \ldots, v_n \), where each consecutive pair \( (v_i, v_{i+1}) \in E \) for \( 1 \leq i < n \). Within this path, \( a \) and \( b \) are referred to as the \textit{head} and \textit{terminus}, respectively, while all other vertices are designated as \textit{intermediate vertices}. The length of a path $P(a, b)$ is $|P(a, b)|$, which denotes the number of vertices on it. The distance from $a$ to $b$ is the length of the shortest path $P(a, b)$ in $G$.

A Gaussian network is defined by a complex modulus $\alpha = a + bi$ (with $a,b \in \mathbb{Z}$) and consists of all Gaussian integers modulo $\alpha$. When $\alpha=3$ (i.e. $3+4i$), there are $|\alpha|^2 = 25$ nodes, labeled by $x+yi$ with $0 \le x,y <5$. Each node has up to four neighbors corresponding to adding or subtracting 1 or $i$ (in the modular ring). The network graph $G=(V, E)$ is implicitly formed by unit step movement, defined by the four direction vectors $\mathbf{d} \in \{GI(1,0), GI(-1,0), GI(0,1), GI(0,-1)\}$. An edge exists between $c$ and $n$ if and only if $n \equiv c + \mathbf{d} \pmod{\alpha}$ for some $\mathbf{d}$. This structure results in a highly regular, fixed-degree (degree-four), toroidal-mesh-like graph. 

\section{Methodology}
\label{section:methadology}
Greedy adaptive routing schemes typically rely on local metrics (such as shortest distance or minimal hop count) to make forwarding decisions. While these approaches are efficient and easy to implement, they suffer from inherent limitations in the presence of faults or dynamic network conditions. Specifically, local decision-making can lead to routing into local minima, where no immediate neighbor appears favorable even though a valid global path exists.

\subsection{Fault Model}
We inject faults by randomly selecting a given percentage of nodes to fail (become unavailable). A faulty node cannot forward packets or be used in a route. Faults are uniformly random, and we vary the fault density (fraction of nodes failed) from 0\% up to 40\% to stress-test the routing. For simplicity, we assume link failures via failed nodes; routing must automatically bypass any failed node.

\subsection{Greedy Adaptive Routing}
The greedy adaptive algorithm forwards each packet at each hop to the neighbor that most reduces the distance to the destination, skipping any neighbor that is failed. Typically, we use the Euclidean distance in the complex plane $|z_{next} - z_{dest}|$. If the “greedy” neighbor is faulty or blocked, the algorithm chooses the best remaining neighbor (adaptive selection). If no neighbor improves the distance, the packet is dropped. Greedy routing requires no learning; it simply uses local information. This method is fast but can become stuck if all closer neighbors are unavailable due to faults \cite{karp2003gpsr}.

\subsection{Reinforcement Learning Routing}
In the RL approach, we treat each node as a state and actions as choosing one of the (up to four) neighboring nodes. The agent’s goal is to learn a policy that directs packets from source to destination.  The adaptive routing strategy is formalized as a Markov Decision Process (MDP) and solved using a Proximal Policy Optimization (PPO) agent, a stability-focused, on-policy Policy Gradient method, well suited for discrete action spaces such as routing decisions. The agent's learning is dependent on a highly specific environment design structured around the network's constraints. 

The environment's State Representation ($s$) utilizes the current node's Gaussian Integer ($GI$) coordinates ($s \in \{a + bi \mid a, b \in \mathbb{Z}\}$). This choice of state is crucial as it leverages the inherent structure of the Gaussian Integer network topology. Unlike simple indexing, this representation preserves the relative spatial and topological relationship between the current node, its neighbors, and the distant destination, providing the PPO agent with contextual information that allows it to learn generalized movement rules and effectively navigate around faults. The Action Space ($a$) is discrete, comprising the four primary directional movements: $a \in \{0, 1, 2, 3\}$ which denotes $a \in \{North, South, East, West\}$, respectively. The Reward Function $R(s, a, s')$ is meticulously engineered to achieve the dual objectives of maximizing path success and minimizing path length under faulty conditions, providing the PPO agent with a significant advantage over the Greedy algorithm:$$R(s, a, s') = \begin{cases}
+100 & \text{if } s' = \text{dst} \\
-50 & \text{if } s' \in \text{Faults} \\
-1 & \text{otherwise (hop cost)}
\end{cases}$$

\begin{itemize}
    \item \textbf{Goal Achievement }($r = +100$): This large positive reward is provided only when the agent reaches the Destination ($\text{dst}$). This sparse, high-magnitude reward strongly incentivizes the agent to complete the path successfully.
    \item \textbf{Fault Collision Penalty} ($r = -50$): This severe negative penalty is applied if the agent attempts to move to a node that is marked as faulty. This is the mechanism by which the agent learns to adaptively avoid damaged network segments, a capability the purely geometric-based Greedy algorithm lacks. The episode is immediately terminated upon receiving this penalty.
    \item \textbf{Hop Cost} ($r = -1$): A small negative reward is given for every successful movement to a non-faulty, non-destination node. This continuous cost encourages the agent to find the shortest possible path, optimizing network efficiency, a crucial factor in the subsequent throughput analysis.

\end{itemize}

The PPO agent operates using an Actor-Critic architecture, where the Policy Network ($\pi_{\theta}$) learns a stochastic mapping from state to action probabilities, and the Value Network ($V_{\phi}$) estimates the expected return $V(s)$. Training is conducted on-policy, collecting a batch of trajectories $\mathcal{D}$ before updating both networks multiple times using the Generalized Advantage Estimation (GAE) and the Clipped Surrogate Objective to stabilize gradient updates. The core of the PPO optimization involves the following steps: Advantage Calculation (GAE): The advantage $\hat{A}_t$ is calculated using the Value Network's temporal difference error $\delta_t$ to reduce variance:$$\delta_t = R_t + \gamma V_{\phi}(s_{t+1}) - V_{\phi}(s_t)$$$$\hat{A}_t = \sum_{l=0}^{T-t} (\gamma\lambda)^l \delta_{t+l}$$where $\gamma=0.95$ is the discount factor and $\lambda=0.92$ is the GAE parameter. Policy Update (Clipped Objective): The Policy Network is updated by maximizing the PPO objective function, which constrains the probability ratio $r_t(\theta)$ between the new policy ($\pi_{\theta}$) and the old policy ($\pi_{\theta_{\text{old}}}$) within a clipping margin $\epsilon=0.2$:$$L^{\text{CLIP}}(\theta) = \hat{\mathbb{E}}_t \left[ \min\left( r_t(\theta) \hat{A}_t, \text{clip}(r_t(\theta), 1-\epsilon, 1+\epsilon) \hat{A}_t \right) \right]$$where $r_t(\theta) = \frac{\pi_{\theta}(a_t|s_t)}{\pi_{\theta_{\text{old}}}(a_t|s_t)}$. The Value Network is simultaneously updated by minimizing the Mean Squared Error (MSE) loss: $L^{\text{VF}}(\phi) = \hat{\mathbb{E}}_t \left[ (R_t - V_{\phi}(s_t))^2 \right]$. The equation $L^{\text{VF}}(\phi) = \hat{\mathbb{E}}_t \left[ (R_t - V_\phi(s_t))^2 \right]$ represents the 
\textbf{Value Function Loss}, where $L^{\text{VF}}(\phi)$ is the objective function being minimized to 
improve the model's accuracy. The term $\phi$ represents the trainable parameters or weights 
of the neural network, while $\hat{\mathbb{E}}_t$ denotes the empirical expectation, or the 
average error calculated over a sampled batch of data at time $t$. Inside the brackets, $R_t$ 
is the target return (the actual accumulated reward the agent received), and $V_\phi(s_t)$ 
is the value predicted by the model for the current state $s_t$. The entire expression 
calculates the \textbf{Mean Squared Error} between the actual and predicted rewards, 
allowing the system to update $\phi$ so that future predictions align more closely with reality.

\section{Experimental Evaluation}
\label{section:experimental}
In this section we carry out a various set of experiments to highlight the performance difference between greedy adaptive and RL-based routing. 

The Greedy Adaptive Routing algorithm utilizes a purely local, deterministic heuristic to navigate the network, prioritizing the minimization of geometric distance to the destination. At every current node $c$, the algorithm evaluates all available neighbors $n$ that are neither faulty nor previously visited. The selection of the next hop is based on the squared Euclidean distance metric (or norm), $D(n, \text{dst})$, where $n$ and $\text{dst}$ are nodes represented by Gaussian Integers, $NB$ is the neighbor nodes, $F$ is the faulty nodes and $V$ is the visited nodes. The router always chooses the neighbor $\hat{n}$ that minimizes this distance:$$\hat{n} = \underset{n \in \text{NB}(c) \setminus (\text{F} \cup \text{V})}{\operatorname{argmin}} D(n, \text{dst})$$. 

This strategy is strictly source-independent and memory-less regarding global topology, relying solely on the local gradient provided by the destination's fixed coordinates. While this approach is computationally efficient (requiring no complex training or look-up tables), it suffers from a critical drawback: it can become trapped in local minima where the current node's path to the destination is blocked (e.g., all neighboring nodes are faulty or lead away from the destination), leading to routing failure without the ability to backtrack or seek alternative, non-geometrically optimal detours.  The path is terminated if the maximum hop count (set as $2 \times N_{\text{NODES}}$ to prevent infinite loops) is reached, or if the set of potential neighbors is empty ($\text{$NB$}(c) \cap \overline{(\text{$F$} \cup \text{$V$})} = \emptyset$), signifying a dead end or trap. The algorithm describing the greedy adaptive routing is shown in Algorithm \ref{alg:greedy_routing}.








\begin{algorithm}[h!]
\caption{Greedy Adaptive Routing}
\label{alg:greedy_routing}
\begin{algorithmic}[1] 
\Require Graph $G$, Source $s$, Destination $d$, Set of Faulty Nodes $F$
\Ensure Path $P$, Success Flag $Success$, Number of Hops $H$

\State $P \leftarrow [s]$ \Comment{Initialize Path with Source}
\State $current \leftarrow s$
\State $Visited \leftarrow \{s\}$
\State $MaxHops \leftarrow 2 \times |Nodes(G)|$
\State $H \leftarrow 0$

\While{$current \neq d$ \textbf{and} $H < MaxHops$}
    \State $N_{potential} \leftarrow \{n \in Neighbors(current) \mid n \notin F \text{ \textbf{and} } n \notin Visited\}$ 

    \If{$N_{potential}$ is empty}
        \State \Return $P, \text{Failure}, -1$ \Comment{Stuck (Local Minimum)}
    \EndIf

    \State $NextNode \leftarrow \arg\min_{n \in N_{potential}} distance(n, d)$
    \State $current \leftarrow NextNode$
    \State Append $current$ to $P$
    \State Add $current$ to $Visited$
    \State $H \leftarrow H + 1$
\EndWhile

\If{$current = d$}
    \State \Return $P, \text{Success}, H$
\Else
    \State \Return $P, \text{Failure}, -1$
\EndIf
\end{algorithmic}
\end{algorithm}

\subsection{RL Agent Training Convergence}
The training progression of the PPO-based RL during the training phase is performed for $N_{episodes}$ (500 episodes in the code) for each fault density trial. The dataset range of values for network parameter $\alpha$ was $2\le \alpha \le 9$ to vary the training dataset of the trained RL. The simulation parameters were fine tuned using Optuna \cite{akiba2019optunanextgenerationhyperparameteroptimization} to select the best set of RL hyper-parameters for evaluating adaptive network protocols. The fine-tuned RL hyper-parameters are shown in Table 1 below.

\begin{table}[h!]
    \centering
    \label{tab:parameters}
    \begin{tabular}{|p{2.5cm}|p{3.5cm}|c|p{5cm}|}
        \hline
        \textbf{Parameter Category} & \textbf{Parameter Name} & \textbf{Value} & \textbf{Rationale/Context} \\
        \hline

        RL Training & Episodes ($RL_{\text{EPISODES}}$) & 500 & Sufficient episodes for PPO convergence in the small state space. \\
        \hline
        RL Hyperparameter & Discount Factor ($\gamma$) & 0.95 & High value to promote long-term policy and the learning of detours. \\
        \hline
        RL Hyperparameter & Learning Rate ($\alpha$) & 0.1 & Standard value for stable RL Learning updates. \\
        \hline
        RL Hyperparameter & Exploration Decay ($\epsilon_{\text{decay}}$) & 0.995 & Gradual shift from exploration to exploitation. \\
        \hline
    \end{tabular}
    \caption{Training Parameters for RL Routing}

\end{table}

\begin{figure}[h!]
    \centering
    \includegraphics[width=0.7\linewidth]{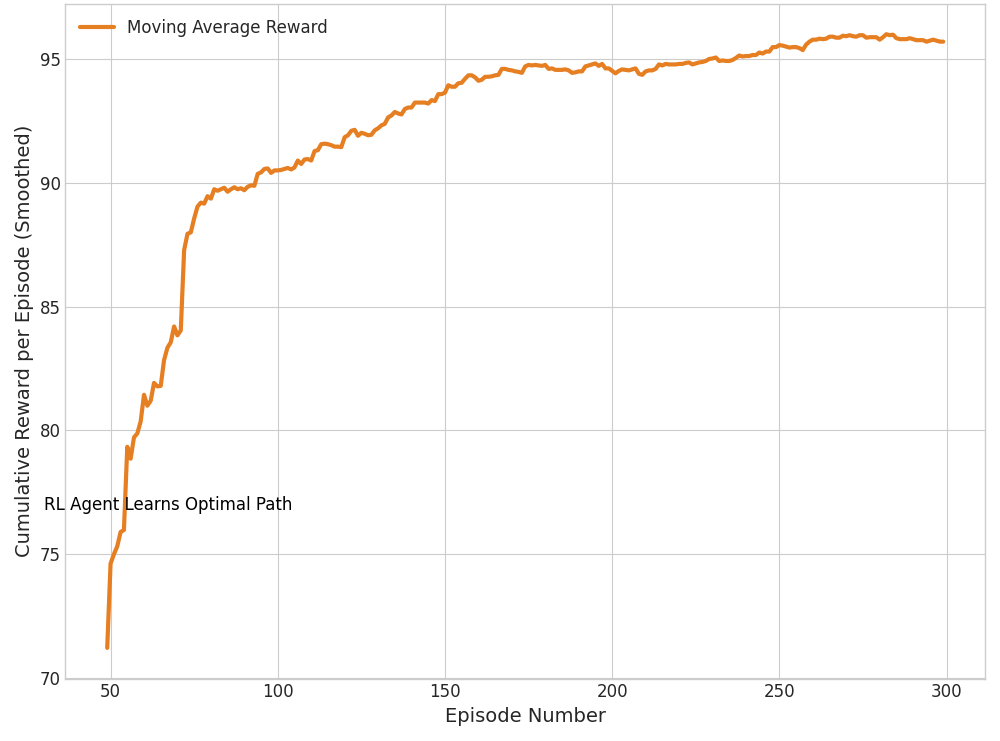}
    \caption{Tuned Hyper-parameters RL average reward} 
    \label{fig:RL_Training_Gaussian}
\end{figure}

The training progression of the RL agent, monitored over $RL_{EPISODES}=500$ episodes, demonstrates a rapid and stable convergence as shown Figure \ref{fig:RL_Training_Gaussian} (represented by the smoothed cumulative reward per episode). The cumulative reward increases sharply within the first 100-150 episodes and stabilizes, indicating that the agent successfully learns the optimal path policy, which is characterized by a balance between the minimum hop cost and the maximum destination reward. This efficiency in policy learning is essential for protocols intended for dynamic deployment. The stability of the final policy ensures that the routing decisions during the inference stage are based on the globally-optimized, long-term expected reward.

\subsection{Faulty Nodes Routing}
\begin{figure}[htbp]
    \centering
    \includegraphics[width=0.7\linewidth]{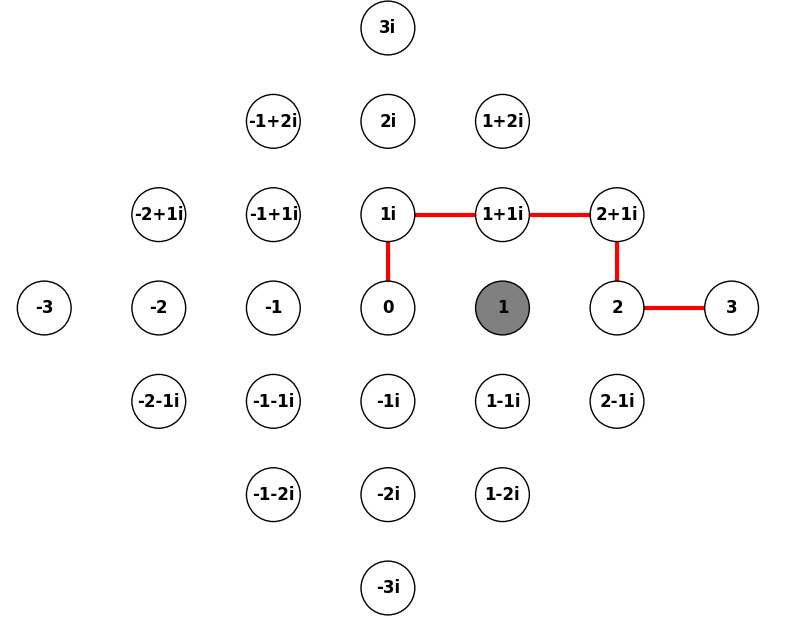}
    \caption{Greedy Adaptive vs RL-based routing with faulty node} 
    \label{fig:GreedyRoutingGaussian}
\end{figure}

\begin{figure}[htbp]
    \centering
    \includegraphics[width=0.7\linewidth]{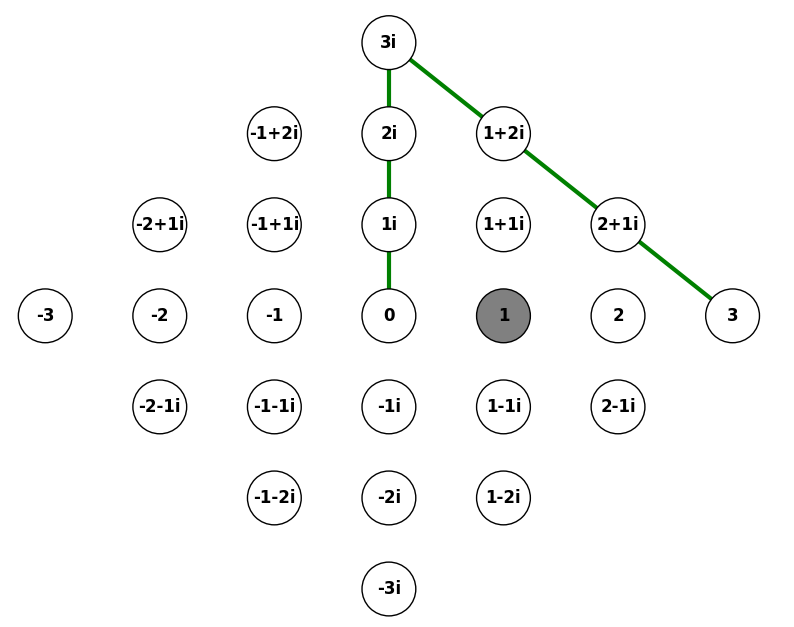}
    \caption{Greedy Adaptive vs RL-based routing with faulty node} 
    \label{fig:RLRoutingGaussian}
\end{figure}

To further clarify the behavioral differences between Greedy Adaptive Routing and the proposed RL-based routing, Figure \ref{fig:GreedyRoutingGaussian} and \ref{fig:RLRoutingGaussian} illustrates a representative routing instance on the Gaussian network with modulus $\alpha=3+4i$. The network consists of 25 nodes labeled by Gaussian integers, with the source node located at 0 and the destination at 3. A single faulty node (shown in gray) is placed near the direct geometric path between source and destination, intentionally creating a challenging routing scenario. In the greedy adaptive case shown in Figure \ref{fig:GreedyRoutingGaussian}, the algorithm selects next hops based solely on local distance minimization. The greedy router initially progresses toward the destination but is forced to detour to node (\textit{i}) once it encounters the faulty region. Because it lacks global awareness or memory, it selects a locally optimal but longer path, resulting in a total of \textbf{five hops} before reaching the destination. Although delivery is successful in this instance, the path is suboptimal and highlights the greedy algorithm’s sensitivity to obstacle placement and its inability to anticipate future constraints. 

In contrast, the RL-based routing agent shown in Figure \ref{fig:RLRoutingGaussian} selects a non-minimal initial move through node (\textit{i}) and then to node (\textit{2i}) that temporarily increases the geometric distance to the destination but ultimately yields a shorter and more reliable route. Through the RL training, the agent has learned to associate specific state–action pairs with long-term success rather than immediate distance reduction. As a result, it deliberately avoids the faulty region by routing through the upper portion of the network and using the wrap around edge from node (\textit{3i}) to node (\textit{3}), reaching the destination in only \textbf{four hops}. This behavior demonstrates a key methodological advantage of the RL approach: routing decisions are driven by cumulative reward optimization rather than greedy heuristics. The RL agent implicitly encodes global structural knowledge of the Gaussian topology and fault patterns within its learned training, enabling it to escape local minima and select paths that appear suboptimal in the short term but are optimal in the long term. This example provides an intuitive explanation for the higher packet delivery ratios and improved throughput observed in the experimental results, particularly under moderate to high fault densities.

\subsection{Faulty Multi-Route Average Distance}
\begin{figure}[h!]
    \centering
    \includegraphics[scale=0.75]{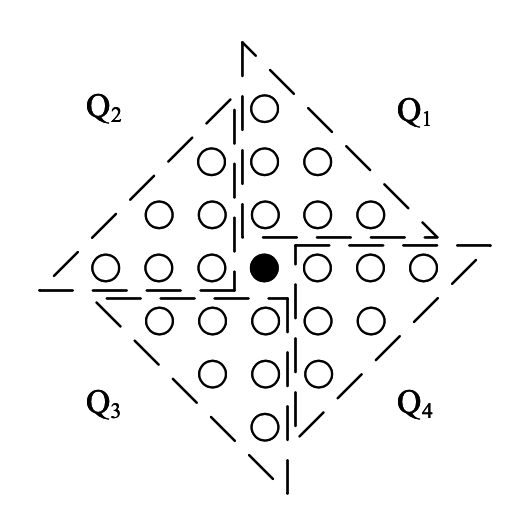}
    \caption{Gaussian Network Quadrants for $\alpha$ = 3 + 4\textit{i}} 
    \label{fig:quadrants}
\end{figure}

We carried out an extended experiment to those shown in Figures \ref{fig:GreedyRoutingGaussian} and \ref{fig:RLRoutingGaussian} which were conducted on the Gaussian network represented by $\alpha = 3+4i$, with a range of multiple faults count. Since the gaussian network is symmetric \cite{Martinez2002Gaussian}, the topology can be divided into 4 equal $Quadrants$ as illustrated in Figure \ref{fig:quadrants}. The layout is partitioned into four distinct quadrants based on the signs of the real and imaginary components. The mathematical distribution of each quadrant is:

\[
Q_1 = \{x+yi \mid x \geq 0, y > 0 \}
\]

\[
Q_2 = \{x+yi \mid x < 0, y \geq 0 \}
\]

\[
Q_3 = \{x+yi \mid x \leq 0, y < 0 \}
\]

\[
Q_4 = \{x+yi \mid x > 0, y \leq 0 \}
\]

This quadrant-based division is mathematically significant due to the rotational symmetry or isotropy of the network; specifically, if a node $z$ exists in $Q_1$, its symmetric counterparts are generated by the operator $i$ such that $iz \in Q_2$, $-z \in Q_3$, and $-iz \in Q_4$. Within a finite network defined by a generator $\alpha$, these quadrants are constrained by a fundamental region or Voronoi cell containing $N = \|\alpha\|^2$ nodes, where the boundary of each quadrant exhibits toroidal wrap-around properties.

\begin{figure}[htbp]
    \centering
    \includegraphics[width=0.7\linewidth]{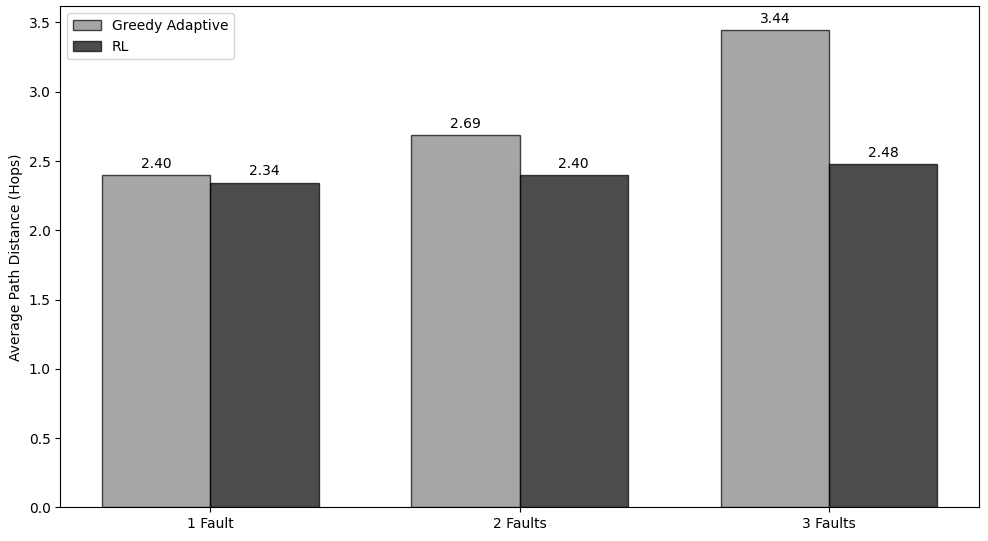}
    \caption{Faulty Multi-Route Average Distance (Quadrant I) for network $\alpha = 3+4i$} 
    \label{fig:faultyMultiRouteAverageDistance}
\end{figure}

We use this $average\ distance$ property to measure the impact of the average routing distance from source node to destination node in a given quadrant in the presence of different faulty nodes count. The experiment reveals a distinct divergence in performance between Greedy Adaptive and Reinforcement Learning (RL) routing within a single quadrant. By isolating the testbed to $Quadrant$ I, the simulation evaluated the efficiency of paths originating from node 0 to all other unique nodes in that quadrant, subjecting the system to a localized concentration of faults. When a single faulty node was introduced, the Greedy Adaptive algorithm achieved an average distance of 2.4 hops, while the RL agent optimized the route to 2.34 hops. As the fault density increased to 2 nodes, the Greedy algorithm’s average distance rose to 2.69, whereas the RL agent remained stable at 2.40. At the maximum stress level of 3 faulty nodes, the Greedy algorithm further jumped to an average distance of 3.44 hops, frequently struggling with the high fault density, while the RL agent maintained a high degree of optimality at 2.48 hops. These results demonstrate that while the Greedy algorithm is highly susceptible to the "local minimum" traps inherent in a congested quadrant, the RL agent leverages the global toroidal symmetry of the Gaussian network to find alternative paths (often wrapping through other quadrants) to maintain near-optimal distance despite the local failures. The demonstration of this average path distance (hops) with the presence of faulty nodes is shown in Figure \ref{fig:faultyMultiRouteAverageDistance}.

\subsection{Packet Delivery Ratio}
The capacity of a routing protocol to maintain high Packet Delivery Ratio (PDR) as the topology degrades is the critical measure of its fault adaptivity. 

\subsubsection{Packet Delivery Ratio (PDR) vs. Fault Density}
\begin{figure}[h!]
    \centering
    \includegraphics[width=0.7\linewidth]{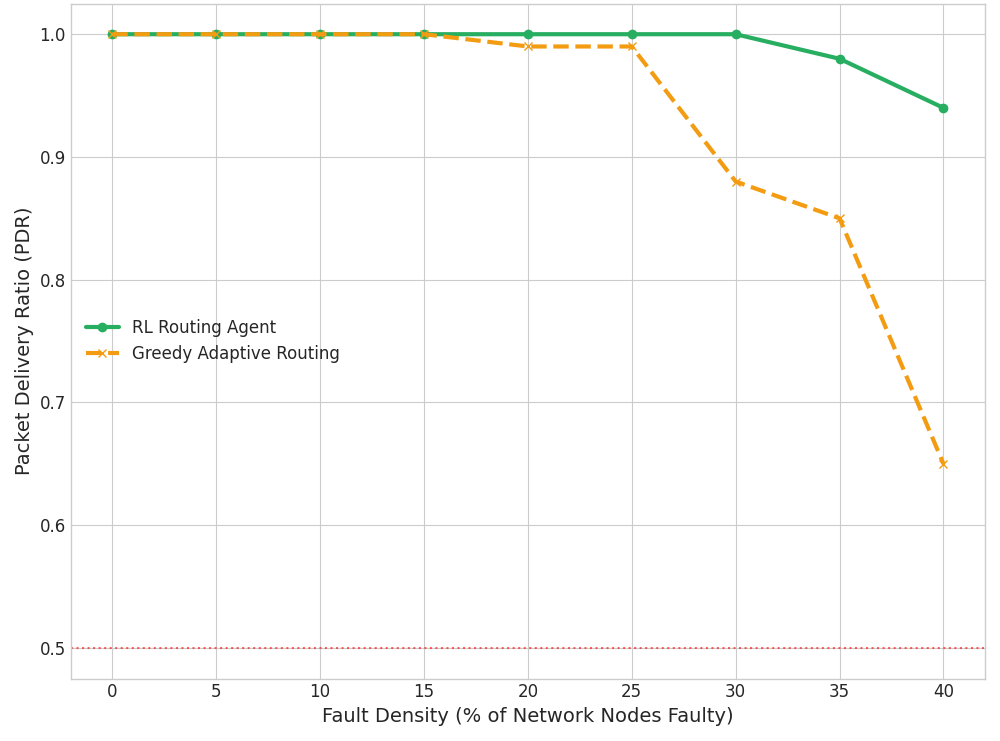}
    \caption{Packet Delivery Ratio (PDR) vs Increased Fault Density} 
    \label{fig:PDRvsFaultDesity}
\end{figure}

The comparison of PDR as a function of increasing fault density provides a clear distinction between the two approaches, as illustrated in Figure \ref{fig:PDRvsFaultDesity}. The results show that the RL-Route agent (solid line) exhibits exceptional resilience, maintaining a perfect or near-perfect $PDR \geq 0.99$ until the fault density approaches $25\%$. Beyond this point, while the performance exhibits a slight decline, the agent still manages to deliver approximately $95\%$ of packets at the extreme $40\%$ fault density. In contrast, the Greedy Adaptive Routing protocol (dashed line) begins to show significant performance degradation starting around the $15-20\%$ fault density mark. By the maximum fault density of $40\%$, the Greedy PDR plummets to approximately $68\%$.

This performance gap is a direct consequence of the two protocols' decision-making philosophies. The Greedy protocol fails because the increasing density exponentially increases the likelihood of creating voids that trap the packet in an unrecoverable local minimum. The RL agent, informed by the negative fault penalty, has learned to take temporary non-minimal steps—a strategy forbidden by the Greedy heuristic—to circumnavigate these fault-induced geometric barriers, thereby exploiting the intrinsic redundancy of the Gaussian network topology.

\subsubsection{Fault Adaptive Score}
\begin{figure}[htbp]
    \centering
    \includegraphics[width=0.7\linewidth]{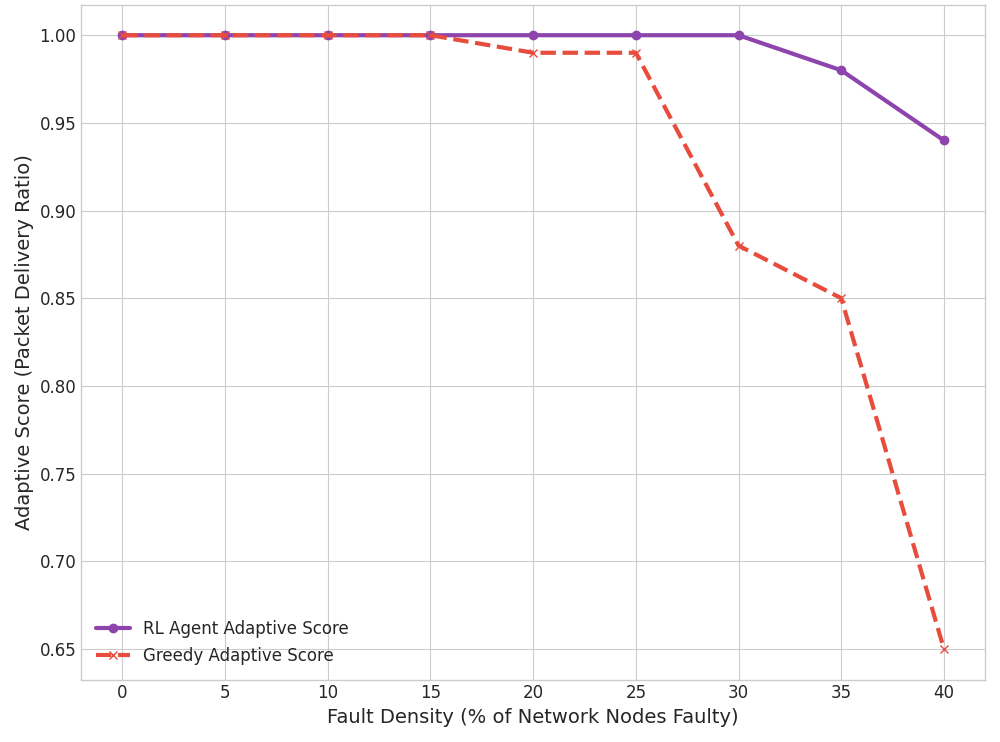}
    \caption{Fault Adaptive Score vs Increased Fault Density} 
    \label{fig:FaultAdaptivePlot}
\end{figure}

Next we compute the Fault Adaptive Score of both approaches and emphasizes the extent of adaptation achieved by each protocol as shown in Figure \ref{fig:FaultAdaptivePlot}.

The magnitude of the RL Agent's illustrates an improved form of network adaptation on the long run. The Greedy method performs a local, metric-based adaptation, which is brittle because it is based on instantaneous geometric improvement. The RL-Route, however, performs policy-based global adaptation. This global policy informs every local decision, allowing the agent to anticipate the failure of a path several hops in advance and select a safer, alternative path, even if it is initially longer. This comprehensive, forward-looking adaptation allows the RL agent to maintain a high Adaptive Score across all fault conditions.

\subsection{Comparative Analysis of Network Efficiency}
While PDR measures reliability, Normalized Throughput (effective delivery rate) measures the efficiency of the successful delivery, especially under network congestion.

\subsubsection{Normalized Throughput vs. Network Load}
\begin{figure}[htbp]
    \centering
    \includegraphics[width=0.7\linewidth]{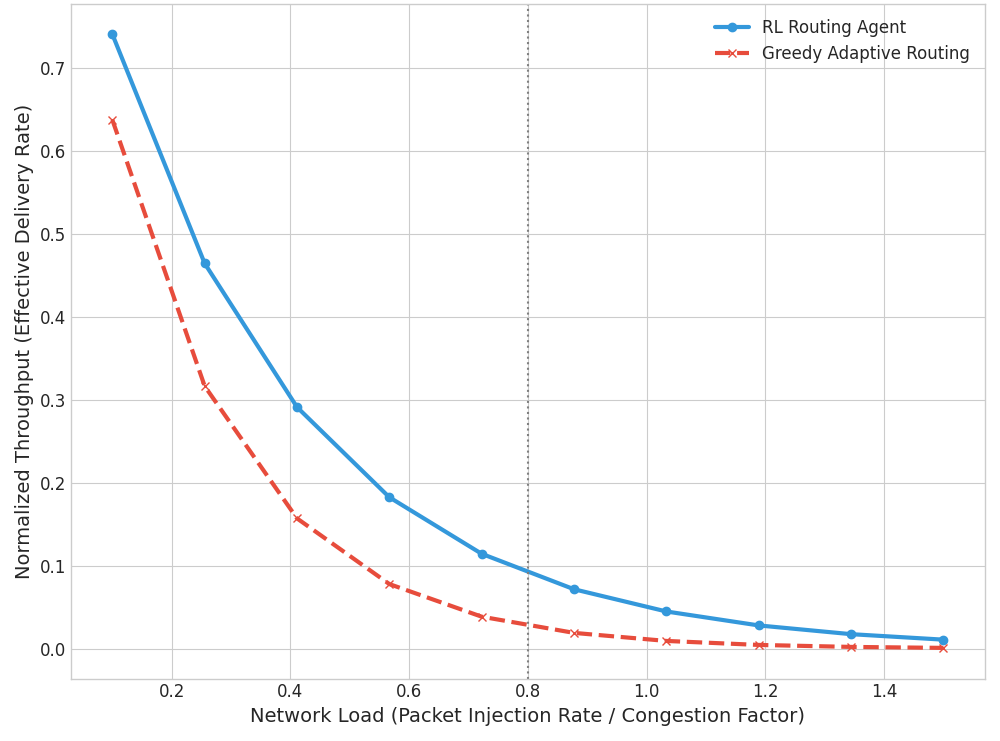}
    \caption{Normalized Throughput vs Network Load} 
    \label{fig:normalizedThroughputPlot}
\end{figure}

An experiment was carried out to compare the Normalized Throughput against Network Load (modeled as Packet Injection Rate/Congestion Factor). As shown in Figure \ref{fig:normalizedThroughputPlot}, the the RL-Route agent (solid line) demonstrates a decisive efficiency advantage against greedy adaptive alogorithm (dotted line). While both approaches tend to behave in a similar trend, at low load factors (e.g., Load $\approx 0.1$), the throughput difference is noticeable, with the RL-Route achieving a throughput of approximately $0.72$ compared to the Greedy’s $0.64$. However, under high network load conditions (e.g., Load $\approx 0.8$, marked by the vertical dashed line), the Greedy throughput approaches zero, indicating near-total collapse under congestion, whereas the RL Agent maintains a small but non-zero throughput.

\subsubsection{Causal Chain for Throughput Improvement}
The robust throughput performance of the RL-routing is a composite function of three factors, all traceable back to its policy design:
\begin{itemize}
    \item \textbf{High PDR (Reliability)}: A packet must be delivered to contribute to throughput. The RL agent’s substantially higher PDR (Figure \ref{fig:PDRvsFaultDesity}) provides a larger baseline of successful deliveries.
    \item \textbf{Path Length Optimization}: Throughput is fundamentally sensitive to path length under congestion. A key aspect of the simulated throughput model is an inverse exponential dependence on both load and path length. The $R=-1$ Step Cost in the RL reward function compels the agent to select the shortest feasible path—the minimum-hop path that also avoids faults. This path is shorter than the non-optimal paths a less-trained agent might take, and crucially, shorter than the non-existent or significantly detoured paths the Greedy agent might find.
    \item \textbf{Fault-Aware Path Selection}: The RL policy implicitly learns to avoid not only faulty nodes but also paths that are likely to be heavily congested or have a high probability of failure. The selection of an optimized, low-hop path ensures a better exponential decay factor under high-load conditions, causing the significant divergence and improved performance. The RL-route, therefore, offers an efficient solution to routing that simultaneously addresses reliability and network efficiency.
\end{itemize}

\subsection{Analysis of Critical Performance Divergence}
The critical divergence in throughput performance in Figure \ref{fig:normalizedThroughputPlot} is most pronounced in the high-load regime, specifically past the Network Load factor of $0.8$. At this point, the Greedy Adaptive Routing's performance decays precipitously toward a throughput of zero. This collapse is not solely due to path inefficiency but is exacerbated by the accumulation of dropped packets (failed PDR) that saturate the network resources before they can be retired. When the Greedy algorithm is trapped, it consumes resources (time, buffer space) until failure, which dramatically reduces overall effective delivery capacity under high injection rates. 

In contrast, the RL-Route maintains a demonstrably higher performance ceiling, achieving a peak normalized throughput of $\approx 0.72$ at the lowest load, significantly exceeding the Greedy protocol's peak of $\approx 0.64$. This initial lead in effective delivery rate confirms that even in non-congested conditions, the RL agent finds inherently shorter paths, fulfilling the path-length minimization mandate of its reward function. Furthermore, the RL-Route's performance curve exhibits a much gentler degradation gradient under increasing load. This resilience stems directly from the robustness observed in its Adaptive Score. Since the RL policy successfully delivers a significantly higher fraction of packets, the overhead of retransmission attempts or path failures is drastically minimized. The successful, short paths learned by the RL agent reduce the total hop count traversed by the average delivered packet, which is essential for maximizing throughput in congestion-sensitive models. Consequently, the RL-Route effectively transforms a fault-adaptive problem (PDR) into a congestion resilience solution (Throughput), affirming the efficacy of its integrated, intelligent decision-making policy in heterogeneous network conditions.

\section{Conclusion}
\label{section:conclusion}
This comparative study rigorously analyzed the fault-adaptivity performance of the PPO based RL routing protocol against a traditional Greedy Adaptive Routing in Gaussian interconnected networks. The experimental results clearly establish the advantage of the RL agent. The Greedy Adaptive Routing protocol, defined by a strict local distance minimization heuristic, proved highly vulnerable to the formation of fault-induced local minima, leading to route incompleteness and a significant degradation of the Packet Delivery Ratio (PDR) under moderate to high fault density. The RL agent, utilizing a multi-objective hybrid reward function that heavily penalizes faulty steps and penalizes long paths, successfully learned a complex, non-geometric policy that facilitates detouring around fault clusters. This policy maintained a high Adaptive Score (PDR) and, through path length minimization, translated its reliability into improved Normalized Throughput across all levels of network load and congestion. The RL policy effectively functions as a Globally-Optimized Local Decision Maker, using long-term value estimation to guide its next hop, a capability in contrast with the purely reactive Greedy method.

Future work will focus on scaling the RL methodology and enhancing its awareness of dynamic network conditions to solidify its application in large-scale systems. Another proposed future work is exploring multi-agent RL for improved router coordination. Further proposed future work includes Hybrid Routing and Heuristic Integration, by dynamically switching between the Greedy and RL. This is beneficial as it seeks to attain the computational speed of heuristics without sacrificing the global optimality and fault-adaptivity capability of the RL model.


\bibliographystyle{elsarticle-num}
\bibliography{ref}

@article{benini2002networks,
  author = {Benini, Luca and De Micheli, Giovanni},
  title = {Networks on Chips: A New {SoC} Paradigm},
  journal = {Computer},
  year = {2002},
  volume = {35},
  number = {1},
  pages = {70--78},
  doi = {10.1109/2.976921},
  issn = {0018-9162},
  publisher = {IEEE}
}

@book{dally2004principles,
  author = {Dally, William J. and Towles, Brian P.},
  title = {Principles and Practices of Interconnection Networks},
  year = {2004},
  publisher = {Morgan Kaufmann Publishers},
  address = {San Francisco, CA},
  isbn = {978-0122007514},
  note = {Comprehensive reference on interconnection network design and analysis}
}

@article{skadron2004temperature,
  author = {Skadron, Kevin and Stan, Mircea R. and Huang, Wei and Velusamy, Sivakumar and Sankaranarayanan, Karthik and Tarjan, David},
  title = {Temperature-Aware Microarchitecture},
  journal = {ACM SIGARCH Computer Architecture News},
  year = {2004},
  volume = {31},
  number = {2},
  pages = {2--13},
  doi = {10.1145/871656.859629},
  publisher = {ACM}
}

@book{sutton2018reinforcement,
  author = {Sutton, Richard S. and Barto, Andrew G.},
  title = {Reinforcement Learning: An Introduction},
  edition = {Second},
  year = {2018},
  publisher = {MIT Press},
  address = {Cambridge, MA},
  isbn = {978-0262039246}
}

@article{kaelbling1996reinforcement,
  author = {Kaelbling, Leslie Pack and Littman, Michael L. and Moore, Andrew W.},
  title = {Reinforcement Learning: A Survey},
  journal = {Journal of Artificial Intelligence Research},
  year = {1996},
  volume = {4},
  pages = {237--285},
  doi = {10.1613/jair.301}
}

@article{mirhoseini2021chip,
  title={A graph placement methodology for fast chip design},
  author={Mirhoseini, Azalia and Goldie, Anna and Yazgan, Mustafa and Jiang, Joe and Songhori, Ehsan and Bae, Soonho and Dabiri, Sajad and Hassabis, Demis and Kavukcuoglu, Koray},
  journal={Nature},
  volume={594},
  number={7862},
  pages={207--212},
  year={2021}
}

@article{Nguyen_2023,
   title={Deep Reinforcement Learning for Cyber Security},
   volume={34},
   ISSN={2162-2388},
   number={8},
   journal={IEEE Transactions on Neural Networks and Learning Systems},
   publisher={Institute of Electrical and Electronics Engineers (IEEE)},
   author={Nguyen, Thanh Thi and Reddi, Vijay Janapa},
   year={2023},
   month=aug, pages={3779–3795} }

@phdthesis{charrwi2025tpu,
  author = {Charrwi, Mohammad Walid M.},
  title = {From TPU to QPU: Bridging Fidelity Gaps Across Next-Generation Computing Systems},
  school = {The City College of New York},
  year = {2025},
  type = {PhD Thesis},
  note = {ProQuest ID: 32237265}
}

@inproceedings{ebda,
  author = {Duato, José},
  title = {EbDa: A New Theory on Design and Verification of Deadlock-free Interconnection Networks},
  booktitle = {Proceedings of the 1997 International Symposium on Computer Architecture},
  year = {1997},
  pages = {1--10},
  doi = {10.1145/264107.264196},
  publisher = {ACM}
}

@ARTICLE{HDGaussian,
author={Bose, Bella and Shamaei, Arash and Flahive, Mary},
journal={ IEEE Transactions on Parallel \& Distributed Systems },
title={{ Higher Dimensional Gaussian Networks }},
year={2016},
volume={27},
number={09},
ISSN={1558-2183},
pages={2628-2638},
doi={10.1109/TPDS.2015.2504936},
publisher={IEEE Computer Society},
address={Los Alamitos, CA, USA},
month=sep}

@techreport{tripathiFaultTolerantRouting,
  author = {Tripathi, Anshuman and Faruqui, Manaal},
  title = {Fault Tolerant Routing in Mesh Topologies},
  institution = {Indian Institute of Technology Kharagpur},
  year = {Year Not Stated in Document},
  type = {Project Report}
}

@article{kurokawa2025adaptive,
  author = {Kurokawa, Yuta and Fukushi, Masaru},
  title = {Adaptive and Passage-Based Fault-Tolerant Routing Methods for Three-Dimensional Mesh {NoCs}},
  journal = {Chips},
  year = {2025},
  volume = {4},
  number = {2},
  article = {14},
  doi = {10.3390/chips4020014}
}

@inproceedings{boyan1993packet,
  author = {Boyan, Justin A. and Littman, Michael L.},
  title = {Packet Routing in Dynamically Changing Networks: A Reinforcement Learning Approach},
  booktitle = {Advances in Neural Information Processing Systems},
  year = {1993},
  volume = {6},
  pages = {671--678},
  publisher = {Morgan Kaufmann},
  note = {NIPS 1993}
}

@article{wu2023rlara,
  author = {Wu, Qiaoyi and Zhang, Zhigang and Li, Yuchun and Wang, Yibo and Pan, Jiliang},
  title = {{RLARA}: A {TSV}-Aware Reinforcement Learning Assisted Fault-Tolerant Routing Algorithm for {3D} {NoC}},
  journal = {Electronics},
  year = {2023},
  volume = {12},
  number = {23},
  article = {4829},
  pages = {1--18},
  doi = {10.3390/electronics12234829},
  publisher = {MDPI}
}

@article{goto2025design,
  author = {Goto, Takeru and Kizumi, Yuki and Iwasaki, Shun},
  title = {Design of Reward Function on Reinforcement Learning for Automated Driving},
  journal = {arXiv preprint},
  year = {2025},
  eprint = {2503.16559},
  archivePrefix = {arXiv},
  primaryClass = {cs.RO}
}

@article{PEDTARA,
author = {Ahmed, Omar and Hu, Min and Ren, Fuji},
year = {2021},
month = {12},
pages = {68},
title = {PEDTARA: Priority-Based Energy Efficient, Delay and Temperature Aware Routing Algorithm Using Multi-Objective Genetic Chaotic Spider Monkey Optimization for Critical Data Transmission in WBANs},
volume = {11},
journal = {Electronics},
doi = {10.3390/electronics11010068}
}

@article{karp2003gpsr,
  author = {Karp, Brad and Kung, H. T.},
  title = {Greedy Perimeter Stateless Routing for Wireless Networks},
  journal = {IEEE/ACM Transactions on Networking},
  year = {2003},
  volume = {11},
  number = {1},
  pages = {96--106},
  month = {Feb},
  doi = {10.1109/TNET.2002.808389},
  issn = {1063-6692}
}

@misc{akiba2019optunanextgenerationhyperparameteroptimization,
      title={Optuna: A Next-generation Hyperparameter Optimization Framework}, 
      author={Takuya Akiba and Shotaro Sano and Toshihiko Yanase and Takeru Ohta and Masanori Koyama},
      year={2019},
      eprint={1907.10902},
      archivePrefix={arXiv},
      primaryClass={cs.LG}
}

@article{Martinez2002Gaussian,
  author={Mart{\'{i}}nez, C. and Beivide, R. and Gabidulin, E. M.},
  journal={IEEE Transactions on Computers}, 
  title={Perfect codes from {G}aussian integers}, 
  year={2002},
  volume={51},
  number={12},
  pages={1456-1461},
  doi={10.1109/TC.2002.1146711},
  note={Discusses the construction of symmetric $G_n$ networks and their fundamental regions.}
}

@inproceedings{elia2003class,
  title={A class of low complexity interconnection networks based on Gaussian integers},
  author={Elia, Michele and Interlando, J Carmelo},
  booktitle={The 11th IEEE International Conference on Networks, 2003. ICON2003.},
  pages={419--423},
  year={2003},
  organization={IEEE},
  note={The foundational paper defining Gaussian Interconnected Networks (GINs) and their structural properties.}
}

@article{huber1994codes,
  title={Codes over Gaussian integers},
  author={Huber, Klaus},
  journal={IEEE Transactions on Information Theory},
  volume={40},
  number={1},
  pages={207--216},
  year={1994},
  publisher={IEEE},
  note={Provides the mathematical background for arithmetic and modulo operations over the ring of Gaussian integers used in network construction.}
}

@book{duato2003interconnection,
  title={Interconnection Networks: An Engineering Approach},
  author={Duato, Jose and Yalamanchili, Sudhakar and Ni, Lionel},
  year={2003},
  publisher={Morgan Kaufmann},
  note={Reference for traditional routing failures, deadlock, and the limitations of static metrics in faulty topologies.}
}

@INPROCEEDINGS{rlquantum,
  author={Charrwi, Mohammad Walid and Ioannou, Georgios and Younis, Ed and de Jong, Wibe Albert and Saeed, Samah Mohamed},
  booktitle={2024 IEEE International Conference on Quantum Computing and Engineering (QCE)}, 
  title={Quantum Circuit Partitioning for Scalable Noise-Aware Quantum Circuit Re-Synthesis}, 
  year={2024},
  volume={02},
  number={},
  pages={359-364},
  doi={10.1109/QCE60285.2024.10306}}

@misc{charrwi2025selfhealingnetworksonchiprldrivenrouting,
      title={Toward Self-Healing Networks-on-Chip: RL-Driven Routing in 2D Torus Architectures}, 
      author={Mohammad Walid Charrwi and Zaid Hussain},
      year={2025},
      eprint={2512.13096},
      archivePrefix={arXiv},
      primaryClass={cs.DC},
      url={https://arxiv.org/abs/2512.13096}, 
}

\end{document}